\documentclass{llncs}

\usepackage{algorithmic}
\usepackage{algorithm}
\usepackage{setspace}

\usepackage{subfigure}
\usepackage{amsmath}
\usepackage{amssymb}
\usepackage{geometry}
\usepackage{float}
\usepackage{graphics}
\usepackage{adjustbox}

\begin{document}

\title{Privacy-Preserving Outsourcing of Large-Scale Nonlinear Programming to the Cloud}

\author{$^\dag$Ang Li\thanks{Ang Li and Wei Du equally contributed to this work. This work was done when Wei Du was at the University of Arkansas.}, $^\S$Wei Du$^\star$, $^\dag$Qinghua Li}
\institute{$^\dag$Department of Computer Science and Computer Engineering, University of Arkansas\\$^\S$Department of Electrical and Computer Engineering, Michigan State University\\
	Email: $^\dag$\{angli,  qinghual\}@uark.edu, $^\S$duwei1@msu.edu}

\maketitle

\begin{abstract}
The increasing massive data generated by various sources has given birth to big data analytics. Solving large-scale nonlinear programming problems (NLPs) is one important big data analytics task that has applications in many domains such as transport and logistics. However, NLPs are usually too computationally expensive for resource-constrained users. Fortunately, cloud computing provides an alternative and economical service for resource-constrained users to outsource their computation tasks to the cloud. However, one major concern with outsourcing NLPs is the leakage of user’s private information contained in NLP formulations and results. Although much work has been done on privacy-preserving outsourcing of computation tasks, little attention has been paid to NLPs. In this paper, we for the first time investigate secure outsourcing of general large-scale NLPs with nonlinear constraints. A secure and efficient transformation scheme at the user side is proposed to protect user’s private information; at the cloud side, generalized reduced gradient method is applied to effectively solve the transformed large-scale NLPs. The proposed protocol is implemented on a cloud computing testbed. Experimental evaluations demonstrate that significant time can be saved for users and the proposed mechanism has the potential for practical use.      
\end{abstract}

\section{Introduction}
Cloud computing has gained an increasing popularity in both academia and industry communities, and been widely used due to its huge computing power, on-demand scalability and low usage cost \cite{murugesan2016encyclopedia}. It offers many services to users, such as data storage, data management and computing resources via the Internet. Besides personal uses such as data storage service represented by Dropbox, cloud computing also has enterprise applications such as big data analytics and business intelligence. One fundamental feature of cloud computing is computation outsourcing, allowing users to perform computations at the resource-rich cloud side and no longer be limited by limited local resources. 

Despite the tremendous benefits, outsourcing computation to the cloud also introduces security and privacy concerns. The first concern is data privacy including both input data privacy and output data privacy \cite{lei2013outsourcing,wang2013harnessing,ren2012security}. The outsourcing paradigm deprives users' direct control over the systems where their data is hosted and computed. The data input to the cloud may contain sensitive information such as medical records and financial asset status. The leakage of these data will breach users' privacy. To protect data against unauthorized leakage, data has to be encrypted before outsourcing. Another concern is the verifiability of results returned from the cloud. Usually users cannot oversee all details of the computations in the cloud. There do exist some motives for the cloud service provider to behave dishonestly and deliver incorrect results to users. One motive is that since intensive computing resources are usually needed to perform outsoursed computation tasks, the cloud service provider might not do all the needed computations to save computing resources. If the cloud server is under outside attacks during the computing process or suffering from internal software failures, the correctness of returned results will also be at risk. Consequently, the verifiability of results returned from cloud should be provided. A third concern is that the computation at the cloud should be efficient; otherwise there is no need for users to outsource computations to the cloud. The time needed by the client to offload the computation to the cloud should be much less than the time needed by the client to solve the computation task by itself \cite{ren2012security,zhou2015outsourcing,chen2014privacy,chen2015new}.

In this paper, we investigate privacy-preserving outsourcing of large-scale NLPs with nonlinear constraints. NLP is a general optimization problem \cite{bazaraa2013nonlinear,bertsekas1999nonlinear}. For instance, finding the optimal investment portfolio is a typical NLP optimization problem subjecting to nonlinear constraints, where an investor wants to maximize expected return and minimize risk simultaneously for investment. In the deep learning area, researchers are always making efforts to find the optimal solution for loss function, which can also be formulated as an NLP with nonlinear constraints \cite{sutskever2013importance}. NLPs with nonlinear constraints are also common in various industry domains, such as the minimum cost of transport and logistics, optimal design, emission-constrained minimum fuel, and so forth \cite{bazaraa2013nonlinear,bertsekas1999nonlinear}. It is very challenging for resource-limited users to solve large-scale NLPs with nonlinear constraints, since it requires intensive computation resources. 

In this work, we propose a privacy-preserving and efficient mechanism to offload large-scale NLPs with nonlinear constraints to the cloud. To the best of our knowledge, privacy-preserving outsourcing of NLPs with nonlinear constraints has never been studied before and this paper is the first. We first formulate the private NLP with nonlinear constraints as a set of matrices and vectors. Then the user generates random vectors and matrices and performs mathematical transformation to protect the original NLP formulation. It is proved that the transformed NLP with nonlinear constraints is computationally indistinguishable from the original one, which means that the cloud cannot infer any useful information about the original NLP from the transformed NLP. At the cloud side, the generalized reduced gradient method is employed to solve the encrypted NLP, which is experimentally demonstrated to be efficient and practical. Finally, the user can verify the correctness of the returned solution to NLP. 

The contributions of this paper can be summarized as follows:
\begin{itemize}
	\item For the first time, we propose an efficient and practical privacy-preserving mechanism for outsourcing large-scale NLPs with nonlinear constraints to the cloud. 
	\item For the proposed solution, we mathematically prove that the input privacy and output privacy of users can be protected. The solution also provides verifiability of cloud-returned results.
	\item The proposed mechanism is implemented, and its performance is evaluated through experiments. The results show high efficiency and practicality of the proposed mechanism.
\end{itemize}

The rest of the paper is organized as follows. Section \ref{sec:related} reviews related work. Section \ref{sec:problem} introduces system model and security definitions. Section \ref{sec:transformation} presents how to use transformation schemes to protect the original NLP and formal proofs are given. Section \ref{sec:algorithm} applies the generalized reduced gradient method to solve the outsourced NLP with nonlinear constraints. Section \ref{sec:evaluation} shows evaluation results. Section \ref{sec:conclusion} concludes this paper.

\section{Related Work}\label{sec:related}
Much work has been done on privacy-preserving outsourcing of computation-intensive tasks to the cloud. Some work focused on outsourcing arbitrary computation functions \cite{kalai2014how,chung2010improved,barbosa2012delegatable}, mainly using fully homomorphic encryption (FHE) schemes such as \cite{gentry2009fully}. Although theoretical guarantees of privacy can be achieved with FHE, current FHE schemes have very high computation cost, making them impractical for large-scale computations such as large-scale NLPs addressed in this paper. Other work designed secure outsourcing protocols for specific problems, such as linear programming \cite{wang2011secure}, system of equations \cite{wang2013harnessing}, distributed linear programming \cite{shen2017distributed}, quadratic programming \cite{zhou2015outsourcing}, and linear regression \cite{chen2013achieving}. Outsourcing basic mathematical computations has also been studied, such as matrix determinant computation \cite{lei2015cloud}, matrix inversion \cite{lei2013outsourcing}, and modular exponentiations \cite{chen2014new}. However, previous work mostly focused on linear systems and some other particular problems. Outsourcing NLPs has received little attention 

Very recently, we also studied securely outsourcing NLPs in our previous work \cite{ducns}, but that work only considers NLPs with \textit{linear} equality constraints. Different from it, this paper addresses NLPs with \textit{nonlinear} inequality and equality constraints which are more complicated and general.

\section{Problem Formulation}\label{sec:problem}
\subsection{NLPs Formulation}
The general form of NLP is expressed as follows \cite{bazaraa2013nonlinear,bertsekas1999nonlinear}:
\begin{eqnarray}
\textbf{P}_1: &\mbox{Minimize} \hspace{0.5cm} &f(\textbf{x}) \nonumber  \\
&\mbox{subject to} \hspace{0.5cm} & \nonumber g_i(\textbf{x}) = 0, \hspace{0.8cm} i = 1,\cdots, m \\ &&  h_j(\textbf{x}) \leq b_j, \hspace{0.6cm} j = 1,\cdots, l \\&& a_k \leq x_k \leq u_k, \hspace{0.1cm} k = 1, \cdots n\nonumber
\end{eqnarray}
where $ \textbf{x} = (x_1, x_2, \cdots, x_n) $ is an $ n $ dimensional vector of variables, $ f(\textbf{x}) $ is a nonlinear objective function, $ g_i(\textbf{x}) = 0 $ are $ m $ equality constraints, and $ h_j(\textbf{x}) \leq b_j $ are $ l $ inequality constraints. In this paper, the NLP is considered as feasible indicating that there exists at least one point $ \textbf{x}^* $ satisfying all of the inequality and equality constraints. Also, it should be noted that the inequality and equality constraints are both of nonlinear form in this paper. NLPs appear many practical applications, such as machine learning, finance budget allocation, and some decision-making problems. Taking the typical support vector machine (SVM) classification as an example. It is known that SVM consists of linear and nonlinear form according to the selection of classification functions. A large portion of the classification tasks require using the nonlinear form of hyperplanes due to the complexity of  data. As a result, the training of the SVM classifier is transformed to solve the nonlinear function subjecting to nonlinear constraints, where nonlinear function is the loss function of SVM model, and nonlinear constraints are nonlinear forms of hyperplanes.

\begin{figure}[h]
	\centering
	\includegraphics[scale=0.55]{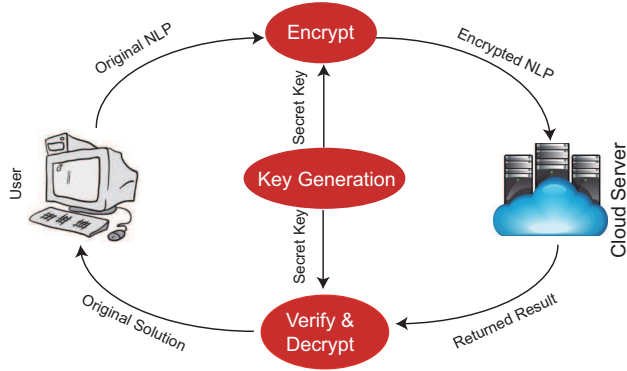}
	\caption{System model for outsourcing large-scale NLPs with nonlinear
		constraints.}\label{fig:issues}
\end{figure}

\subsection{System Model}
The outsourcing model has two parties, the user and the cloud server, as illustrated in Fig. 1. The user has an NLP problem to solve. However, the user cannot solve this large-scale problem due to his limited computation power. Thus, he outsources the NLP problem to the cloud server. In order to protect the original NLP problem $ \Phi $ from being known to the cloud, the user generates a private key $ K $ to encrypt the problem $ \Phi $, and sends the encrypted NLP problem $\Phi(K)$ to the cloud server. The cloud server solves $\Phi(K)$ using the generalized reduced gradient method, and returns the solution back to the user. During the computing process, the cloud server is supposed to learn nothing or very little information about the original NLP problem. When the user receives the solution for $\Phi(K)$ from the server, the user verifies its correctness. If it is incorrect, the user will reject it; if it is correct, the user will accept it and decrypt it with the private key $ K $ to get the solution for the original NLP problem $ \Phi $. 

\subsection{Security Model and Goal}
The security concerns and threats are mainly from the untrusted cloud server. A malicious cloud server may try to learn about the original NLP problem. It may also not follow the correct computing process of the problem and derive a wrong solution. As a result, the security goal is two-fold: hiding the original NLP problem from the cloud in order to protect the user’s privacy, and providing a verification mechanism for the user to check the correctness of returned result so that the cloud server cannot cheat. 


\subsection{Security Requirements}
This section gives a formal security definition for the outsourcing protocol. Let us first look at the scope of private information within this context. In the original NLP problem $ \textbf{P}_1 $, the coefficient matrices of the equality and inequality constraints contain sensitive information. The positions of elements in the coefficient matrices may also contain private information, e.g. the node distribution graph of an optimal digital circuit layout path. In addition, the solution $ \textbf{x}^* $ of the original NLP problem $ \textbf{P}_1 $ should also be protected. 

The concept of computational indistinguishability is used in this paper to design a secure outsourcing protocol. 

\textit{Definition 1}: A randomized algorithm $ \mathcal{A} $ satisfies computational indistinguishability if and only if for any two databases $ D $ and $ D' $, for every probabilistic polynomial-time adversary machine \textit{M}, there exists a negligible function \textit{$neg(\cdot)$} such that \cite{katz2014introduction}:
\begin{equation}
|Pr[\textit{M}^\mathcal{A}(D)] - Pr[\textit{M}^\mathcal{A}(D')]| \leq neg(\cdot)
\end{equation}
where the notation $\textit{M}^\mathcal{A}(D)$ (similarly for $ \textit{M}^\mathcal{A}(D') $) means that adversary machines have access to the database and try to extract private information from the data. \textit{Definition 1} measures the information leakage level of the encryption scheme that encrypts the original NLP problem. If computational indistinguishability is achieved, the cloud server cannot learn anything significant about the original NLP problem.

\section{NLP Transformation}\label{sec:transformation}
\subsection{Input Privacy Protection}
In order to protect the coefficient matrices and vectors of the constraints as shown in the general form of NLP $ \textbf{P}_1 $, they are encrypted by the user’s privacy key $ K $.

\subsubsection{Protecting Equality Constraints}
Suppose the coefficient matrix and the vector of the equality constraints in $ \textbf{P}_1 $ are denoted as $ \textbf{G} \in \mathbb{R}^{m \times n}$ and $ \textbf{b} \in \mathbb{R}^{m \times 1} $, respectively. $\textbf{G} $ and $ \textbf{b} $ can be efficiently hidden by employing matrix multiplications. In particular, the user can protect the equality constraint matrix and vectors as follows:
\begin{equation}
\begin{split}
&\hat{\textbf{G}} = \textbf{PQG} \\
&\hat{\textbf{b}} = \hspace{0.1cm}\textbf{PQb}
\end{split}
\end{equation} 
where $ \textbf{P} \in \mathbb{R}^{m \times m}$  is a diagonal matrix, with the elements defined as follows:
\begin{equation}
\textbf{P}_{i,j}=\left\{
\begin{aligned}
&r_i  \hspace{0.8cm} i = j \\
&0  \hspace{0.9cm}   i \neq j \\
\end{aligned}
\right.
\end{equation} 
Here the value of $r_i$ comes from the uniform distribution defined as:
\begin{equation}
r_i=\left\{
\begin{aligned}
&r  \hspace{0.8cm} -N < r < N \\
&0  \hspace{1.0cm} \text{otherwise} \\
\end{aligned}
\right.
\end{equation}
$ \textbf{Q} \in \mathbb{R}^{m \times m} in Eq. (3) $ is a positive constant diagonal matrix, which is expressed as:
\begin{equation}
\textbf{Q}_{i,j}=\left\{
\begin{aligned}
&C  \hspace{0.8cm} i = j \\
&0  \hspace{0.9cm}   i \neq j \\
\end{aligned}
\right.
\end{equation} 
It can be seen from Eq. (3) that the matrices $ \textbf{G} $ and $ \textbf{b} $ are masked by the multiplying a random diagonal matrix $ \textbf{P} $ and a constant diagonal matrix $\textbf{Q}$. It should be noted that the rank of matrix $ \textbf{G} $ stays the same due to the full rank of $ \textbf{P} $ and $ \textbf{Q} $. From the encryption form, one cannot extract any useful information without releasing $ \textbf{P} $ and $ \textbf{Q} $. We will give detailed mathematical analysis of the above transformation in the next section.

\subsubsection{Inequality Constraints Protection}
The coefficient matrix of the inequality constraints is denoted as $ \textbf{H} \in \mathbb{R}^{l \times n}$. A similar approach can be used to encrypt $ \textbf{H} $ as follows:
\begin{equation}
\hat {\textbf{H}} = \textbf{TSH}
\end{equation}
where $ \textbf{T} \in \mathbb{R}^{l \times l}$ is a diagonal matrix with elements generated following the uniform distribution defined in Eq. (5) and $ \textbf{S} \in \mathbb{R}^{l \times l}$ is set to be a diagonal constant matrix. 

\subsection{Output Privacy Protection}
The above matrix multiplication mechanism is able to protect the input privacy, but the output privacy has not been addressed yet. The user sends the NLP to the cloud server, and will receive a solution $\textbf{x}^*$ back from the cloud. The result $\textbf{x}^*$ might contain sensitive information, e.g., the asset allocation strategy in a financial company. In fact, the output privacy can be easily protected by vector addition. We can just replace the original variable vector $\textbf{x}$ with the following:
\begin{equation}
\textbf{z} = \textbf{x} + \textbf{r}
\end{equation}
where the elements of vector $ \textbf{r} $ are also taken from the uniform distribution defined in Eq. (5). After completing the protection of the input and output privacy, we can rewrite the original problem $ \textbf{P}_1 $ as the following:
\begin{eqnarray}
\textbf{P}_2: &\mbox{Minimize} \hspace{0.5cm} &f(\textbf{z}) \nonumber  \\
&\mbox{subject to} \hspace{0.5cm} & \nonumber \hat g_i(\textbf{z}) = 0, \hspace{0.8cm} i = 1,\cdots, m \\ &&  \hat h_j(\textbf{z}) \leq \hat b_j, \hspace{0.6cm} j = 1,\cdots, l \\&& \hat a_k \leq z_k \leq \hat u_k, \hspace{0.1cm}k = 1, \cdots n\nonumber
\end{eqnarray}
where $ \hat{\textbf{G}} = \textbf{PQG} $, $  \hat{\textbf{b}} = \textbf{PQb} $, $ \hat {\textbf{H}} = \textbf{TSH} $, $ \textbf{z} = \textbf{x} + \textbf{r} $, $\hat a_k = a_k + r_k$, and $\hat u_k = u_k + r_k$. 

\subsection{Structure Privacy Protection}
The above matrix transformations can protect the element values within the input and output matrix; however, the structure of the input and output matrix (i.e., positions of non-zero elements) still needs to be protected, which might also contain sensitive information. For example, 　the circuit layout is of vital importance in the area of electronics design, and one of the common methods is to construct matrices according to the node distribution. Thus, it is easy to recover the original circuit layout if we know the circuit matrices. As a result, the position of the elements in a matrix sometimes will contain sensitive and valuable information which needs to be hidden.  
Next we will introduce a matrix permutation mechanism to protect the position information of the original matrix. 

\begin{algorithm}[h]
	\caption{ Key generation.}   
	\label{alg:Framwork}   
	\begin{algorithmic}[1] 
		\REQUIRE ~~\\ 
		Input size $ n $; \\
		\ENSURE ~~\\ 
		Random uniformly distributed vector $ \textbf{r} $; \\
		Random uniformly distributed matrix $ \textbf{S} $; \\
		Constant matrix $ \textbf{M} $; \\
		Random permutation matrix $ \textbf{W} $;\\  
		\STATE Generate a uniformly distributed random vector $ \textbf{r} $ according to Eq. (5);
		\STATE Generate uniformly distributed diagonal matrices $ \textbf{S} $ according to Eq. (5);
		\STATE Generate constant diagonal matrix $ \textbf{M}  $ according to Eq. (6);
		\STATE Set $S = \{ 1, 2, 3, \cdots, n \};$   
		\label{ code:fram:extract }
		\STATE \hspace{0.4cm}\textbf{for} $ j = 1 $ to $ n $ \\
		\label{code:fram:xx}  
		\STATE \hspace{0.8cm} select $ i $ randomly from $ i \in (1,j) $; 
		\STATE \hspace{0.8cm} swap  $ S (i) $ and $ S (j) $;
		\label{code:fram:trainbase}  
		\STATE \hspace{0.3cm} end \textbf{for}   
		\label{code:fram:add}  
		\STATE \hspace{0.4cm}\textbf{for} $ i = 1 $ to $ n $ \\ 
		\STATE \hspace{0.8cm}\textbf{for} $ j = 1 $ to $ n $ \\
		\STATE \hspace{1.2cm} $ \varphi$ outputs the $ i $th element from $ S $ with $ \varphi(i)$;
		\STATE \hspace{1.2cm} $ \sigma$ outputs value with $ \sigma_{\varphi(i),j} $;
		\STATE \hspace{1.2cm} set $ W(i,j) $ = $ \sigma_{\varphi(i),j} $;
		\STATE  \hspace{0.7cm} end \textbf{for} 
		\STATE  \hspace{0.3cm} end \textbf{for} 
		\label{code:fram:select}  
		\RETURN $ \textbf{r} $, $ \textbf{S} $,  $ \textbf{M} $, $\textbf{W} $; 
	\end{algorithmic}  
\end{algorithm}  

The permutation of a matrix starts from permuting a set $ S $. Consider a two line notation representing an original set $ S $ and its permutation set $ S' $ denoted as [2]:
\begin{equation}
\left\{
\begin{aligned}
s_1, s_2, s_3 \cdots, s_n  \\
s'_1, s'_2, s'_3 \cdots, s'_n \\
\end{aligned}
\right\}
\end{equation}
Here the upper line is the elements of the original set $ S $ and the bottom line is the elements from the permutation set $ S' $. Note that $ S $ and $ S' $ have the same elements but with different orders. Here we define $ s'_i = \theta(s_i) $ to represent the function mapping from Eq. (10), indicating that for an upper line element $s_i$ as the input, the output will be the corresponding element $s’_i$ in the bottom line. Then the permutation matrix can be obtained as follows:
\begin{equation}
M(i,j) = \sigma_{\theta(i),j} \hspace{0.6cm} 1 \leq i, j \leq n
\end{equation}
where $\sigma_{i,j}$ is the Kronecker delta function, which is commonly used in engineering field and defined as [5] :
\begin{equation}
\sigma_{i,j}=\left\{
\begin{aligned}
&0  \hspace{0.8cm} i \ne j \\
&1  \hspace{0.8cm} i = j \\
\end{aligned}
\right.
\end{equation}  
and $ \theta(i) $ is defined as above, outputting the $ i $th element of the permutation set.
Then we can protect the position information of matrix $ \hat {\textbf{G}} $ and $ \hat {\textbf{H}} $ with the following expression:
\begin{equation}
\begin{split}
 \textbf{G}' = \textbf{X}\hat {\textbf{G}}\textbf{Y} \\
\textbf{H}' = \textbf{X}\hat {\textbf{H}}\textbf{Y}
\end{split}
\end{equation}
where $ \textbf{X} $ and $ \textbf{Y} $ are random permutation matrices generated from Eq. (11). It can be seen that matrices $ \textbf{X} $ and $ \textbf{Y} $ are used to randomly permute the positions of rows and columns of the matrix, respectively. Since the permutation matrices are randomly generated, they will permute the original matrix to random-order rows and columns. As a result, the cloud server will not be able to learn any structure information from the reordered matrices. 

Up to now, we have finished transforming the original NLP, and the problem $ \textbf{P}_3 $ can be rewritten as:
\begin{eqnarray}
\textbf{P}_3: &\mbox{Minimize} \hspace{0.5cm} &f(\textbf{z}) \nonumber  \\
&\mbox{subject to} \hspace{0.5cm} & \nonumber  g'_i(\textbf{z}) = 0, \hspace{0.8cm} i = 1,\cdots, m \\ &&  h'_j(\textbf{z}) \leq  b_j', \hspace{0.6cm} j = 1,\cdots, l \\&& \hat a_k \leq z_k \leq \hat u_k, \hspace{0.1cm}k = 1, \cdots n\nonumber
\end{eqnarray}
where $ \textbf{G}' = \textbf{X}\hat {\textbf{G}}\textbf{Y} $, $ \textbf{H}' = \textbf{X}\hat {\textbf{H}}\textbf{Y} $, and $ \textbf{b}' = \textbf{X} \hat {\textbf{b}}$. 

Both key generation and matrix transformation are performed by the user locally, and the procedures are summarized in \textit{Algorithm 1} and \textit{Algorithm 2}, respectively.  

\begin{algorithm}[h]   
	\caption{ Transformation mechanism.}   
	\label{alg:Framwork}   
	\begin{algorithmic}[1] 
		\REQUIRE ~~\\ 
		Objective function $f(\textbf{x})$;\\  
		Equality coefficient matrix $\textbf{G}$ and inequality coefficient vector $\textbf{b}$; \\
		Inequality coefficient matrix $\textbf{H}$; \\
		\ENSURE ~~\\ 
		Encrypted objective function $f(\textbf{z})$;\\ 
		Encrypted matrix $\textbf{G}^\prime$ and vector $\textbf{b}^\prime$;\\
		Encrypted matrix $\textbf{H}^\prime$;	
		\STATE Generate a random vector $\textbf{r}$ from \textit{Algorithm 1} to obtain $ \textbf{z} = \textbf{x} + \textbf{r} $, $f(\textbf{z})=f(\textbf{x} + \textbf{r})$; 
		\label{ code:fram:extract }%
		\STATE Generate two random diagonal matrices $\textbf{P}$ and $ \textbf{T} $ from \textit{Algorithm 1};\\
		\STATE Generate two constant diagonal matrices $ \textbf{Q} $ and $ \textbf{S} $ from \textit{Algorithm 1};
		\label{code:fram:xx}  
		\STATE Calculate $ \hat{\mathbf{\textbf{G}}} = \textbf{PQG} $ and $ \hat{\mathbf{\textbf{b}}} = \textbf{PQb} $;
		\STATE Calculate  $ \hat{\mathbf{\textbf{H}}} = \textbf{TSH} $;
		\label{code:fram:trainbase}  
		\STATE Generate matrices $ \textbf{X} $ and $ \textbf{Y} $ from line 4 to 15 in \textit{Algorithm 1} , corresponding to matrix $ \textbf{W} $; 
		\label{code:fram:add}  
		\STATE  Calculate $\textbf{G}\textquoteright = \textbf{X}\hat{\mathbf{G}}\textbf{Y}$ and $ \textbf{b}\textquoteright = \textbf{X}\hat{\mathbf{\textbf{b}}} $;  
		\label{code:fram:ad}  
		\STATE  Calculate $\textbf{H}\textquoteright = \textbf{X}\hat{\mathbf{H}}\textbf{Y}$; 
		\label{code:fram:select}  
		\RETURN $f(\textbf{z}),\textbf{G}\textquoteright, \textbf{b}\textquoteright, \textbf{H}\textquoteright$;
	\end{algorithmic}  
\end{algorithm}  

\subsection{Privacy Analysis}
In order to show why the aforementioned transformation schemes can protect input privacy and output privacy, next we will derive a theorem proving that the input matrix and the output vector are computationally indistinguishable from a randomly generated matrix and vector, respectively. 

\textbf{Theorem 1}. Let the elements of $ \textbf{R} \in \mathbb{R}^{m \times n}$ and $ \textbf{r} \in \mathbb{R}^{n \times 1}$ be generated from the uniform distribution defined in Eq. (5). Then the matrices  $ \hat{\textbf{G}} = \textbf{PQG} $ and  $ \hat{\textbf{H}} = \textbf{TSH} $ are computationally indistinguishable from a random matrix $ \textbf{R} $, and vector $ \textbf{z} = \textbf{x} + \textbf{r} $ is computationally indistinguishable from a random vector $ \textbf{r} $. 

\textit{Proof}: 
Firstly, to prove the computational indistinguishability between matrices $ \hat {\textbf{G}} $ and $ {\textbf{R}} $, we need to show for any probabilistic polynomial-time adversary machines $ \textit{M} $ having access to database $ \textit{M}^\mathcal{A} $, it can only tell the difference between $  {\hat G_{i,j}} $ and $ {R_{i,j}} $ with negligible success probability, where   $  {\hat G_{i,j}} $ is the element in the $i$th row and $j$th column of $\hat {\textbf{G}}$ and $R_{i,j}$ is the element in the $i$th row and $j$th column of $\textbf{R}$. Suppose the adversary machine $ \textit{M} $ sending inquiry to database $ D $, it will output $ Pr[\textit{M}^\mathcal{A}(D)] $, which is the probability of the element coming from a specific database $ D $. It is obvious that if $ M $ determines the element coming from a specific database $ D $ with full confidence, it will output 1. Suppose the element ${\hat G_{i,j}} $ is chosen from $\hat {\textbf{G}}$ by adversary machine $ M $, the success probability to identify it from $\hat {\textbf{G}}$ is expressed as the following:
\begin{equation}
\begin{split}
Pr[\textit{M}^\mathcal{A}(\hat G_{i,j})] &= \frac{1}{2}Pr[-N < \hat G_{i,j} <N]\\
&\hspace{0.4cm}+ Pr[\hat G_{i,j} \leq -N] + Pr[\hat G_{i,j} \geq N]
\\&=\frac{1}{2}[1 - Pr[\hat G_{i,j} \leq -N]] - Pr[\hat G_{i,j} \geq N]]
\\&\hspace{0.4cm}+Pr[\hat G_{i,j} \leq -N] + Pr[\hat G_{i,j} \geq N]
\end{split}
\end{equation}
Here is brief explanation of the above expression, if the inquiry ${\hat G_{i,j}} $ by $ M $ is within the range $ (-N, N) $, the probability of it coming from  $\hat {\textbf{G}}$ is  1/2 since it is possible for both $ \textbf{R} $ and $\hat {\textbf{G}}$ owing elements falling in the range $ (-N, N) $. However, if the inquiry ${\hat G_{i,j}} $ by $ M $ is out of the range $ (-N, N) $, it must be from matrix ${\hat G_{i,j}} $ that the probability is 1. To calculate Eq. (15), we first have that 
\begin{equation}
\begin{split}
	Pr[\hat G_{i,j} \geq N] &= Pr[Q_{i,i}P_{i,i}G_{i,j} \geq N]
	\\&= Pr[P_{i,i}G_{i,j} \geq \frac{N}{C}]
	\\&= \alpha Pr[P_{i,i} \geq \frac{N}{CG_{i,j}}] + (1 - \alpha) Pr[P_{i,i} \leq \frac{N}{CG_{i,j}}]
	\\&\leq \alpha Pr[P_{i,i} \geq \frac{N}{CL}] + (1 - \alpha) Pr[P_{i,i} \leq \frac{-N}{CL}]
	\\& = 1 - \frac{1}{CL}
	\end{split}
\end{equation}
where $ P_{i,i} $ and $Q_{i,i}$ is the element in $ i $th row and $ i $th column of matrix $ \textbf{P} $ and $ \textbf{Q} $, respectively; $ L $ is the maximum value of elements in $ {\hat {\textbf{G}}} $, $ C $ is the constant in $ \textbf{Q} $ defined in Eq. (6), $\alpha$ is the probability for the element $G_{i,j}$ to be positive and $1- \alpha$ is the probability for the element $G_{i,j}$ to be negative. 
\\ \indent In addition, we can similarly obtain :
\begin{equation}
 Pr[\hat G_{i,j} \leq -N] = 1 - \frac{1}{CL}
\end{equation}

Thus, the success probability for adversary machine $M$ determining $\hat G_{i,j}$ chosen from matrix $\hat{\mathbf{G}}$ in Eq. (15) will be:
\begin{equation}
Pr[\textit{M}^\mathcal{A}(\hat G_{i,j})] \leq \frac{3}{2} - \frac{1}{CL}
\end{equation}
The success probability for $R_{i,j}$ to be identified by the distinguisher $M$ is expressed as follows :
\begin{equation}
Pr[\textit{M}^\mathcal{A}(\hat R_{i,j})] =  \frac{1}{2}
\end{equation}
The comparison of Eq. (18) and Eq. (19) will lead to
\begin{equation}
|Pr[\textit{M}^\mathcal{A}(\hat G_{i,j})] - Pr[\textit{M}^\mathcal{A}(\hat R_{i,j})]| \leq  \frac{CL-1}{CL}
\end{equation}
By comparing Eq. (20) and Eq. (2), we can define 
\begin{equation}
neg(C) = \frac{CL-1}{CL}
\end{equation}
Since we can choose any positive constant $ C $, we can choose a value of $C$ that makes $CL$ be close to 1, which will make Eq. (21) be a negligible value. As such, the encrypted matrix $\hat{\textbf{G}}$ and random generated $\textbf{R}$ meet the property of computational indistinguishability. Similarly, we can also prove the computational indistinguishability between $\hat{\textbf{H}}$, ${\textbf{R}}$, and $\textbf{z}$, $\textbf{r}$, respectively. The proof is complete.

It can be concluded from \textbf{Theorem 1} that even if the adversary machines have full access to the data, it still cannot learn any useful information. As a result, sending the encrypted data via the transformation mechanism to the cloud will not release any private information, proving the security of the proposed protocol.

\section{Secure Outsourcing Algorithm For Encrypted NLPs}\label{sec:algorithm}
In this section, we will design an efficient outsourcing algorithm to solve the encrypted large-scale NLPs. To solve large-scale NLPs, the generalized reduced gradient (GRG) method \cite{bazaraa2013nonlinear,ducns} is employed to get the optimal solution of $ \textbf{P}_3 $. The strategy of GRG is based on an iterative way to repeatedly generate feasible improving directions optimizing NLPs.
\subsection{Gradient Decent Method for Unconstrained NLPs}
Before delving into details of the GRG to solve large-scale NLPs subjecting to a system of constraints, we first introduce how to find the optimal solution of an unconstrained objective function. A popular and widely used algorithm for solving unconstrained problem is gradient decent method. Suppose $ f(\textbf{z}) $ is convex and differentiable within the neighborhood of $ \textbf{z}_0 $. The decent method is to produce a sequence of $ \textbf{z}_k $ ($ k = 0, 1, 2, ... $) which can continuously decrease the objective function, expressed as: 
\begin{equation}
\textbf{z}_{k + 1} = \textbf{z}_k + \lambda \textbf{d}
\end{equation}
where $ \textbf{d}$ is called search direction, $ k = 0, 1, 2, ... $ is iteration number, and $ \lambda $ is termed as step length. 
\\ \indent The decent method means that for every iteration of the algorithm, we must have 
\begin{equation}
f(\textbf{z}_{k + 1}) < f(\textbf{z}_k)
\end{equation}
except when $ \textbf{z}_k $ is already an optimal solution of the objective function. 
\\ \indent The convexity of the objective function indicates that the search direction $ \textbf{d} $ must satisfy the following expression:
\begin{equation}
\nabla f(\textbf{z}_k)\textbf{d} < 0
\end{equation}  
It can be seen from Eq.(24) that search direction $ \textbf{d} $  must form an acute angle with the negative gradient, thus as such it is called as a decent direction. As a result, an obvious choice for $ \textbf{d} $ is along the negative gradient direction $ -\nabla f(\textbf{z}_k) $. Once the selection of the search direction is completed, next step is to determine step size as the following:
\begin{equation}
\lambda = \text{arg min}_{s\geq 0} \hspace{0.1cm} f(\textbf{z} + s\textbf{d})
\end{equation}
An exact line search method can be used to solve the one variable optimization task, just as Eq. (25). However, the above gradient decent method cannot be applied to $ \textbf{P}_3 $ due to the existence of constraints. The reason is that if we directly move $ \textbf{z}_k $ along the negative gradient direction $ \textbf{z}_{k + 1} = \textbf{z}_k -\lambda \nabla f(\textbf{z}_k) $, the feasibility of the constraints may be destroyed. As a result, it occurs to us that we have to figure out a way to generate a series of feasible directions gradually approaching the optimal solution of the constrained large-scale NLPs, which will be shown as next section.

\subsection{Generalized Reduced Gradient Method for Constrained NLPs}
GRG method is robust and efficient in solving large-scale nonlinear problems practically. The constraints in $ \textbf{P}_3 $ includes both equality and inequality equations. In fact, we can make all of the inequality constraints $ \hat h_j(\textbf{z}) \leq \hat b_j, j = 1,\cdots, l $ to equality constraints by introducing a bunch of slack variables as follows:
\begin{equation}
\begin{split}
&\hat h_j(\textbf{z}) + s_j - \hat b_j = 0, \hspace{0.2cm} j = 1,\cdots, l
\\ &s_j \geq 0,  \hspace{2.1cm} j = 1,\cdots, l
\end{split}
\end{equation}
Thus we can rewrite $ \textbf{P}_3 $ in the following general form:
\begin{eqnarray}
\textbf{P}_4: &\mbox{Minimize} \hspace{0.1cm} &f(\textbf{z, s}) \nonumber  \\
&\mbox{subject to} \hspace{0.1cm} & \nonumber \hat e_i(\textbf{z, s}) = 0, \hspace{0.4cm} i = 1,\cdots, m, \cdots, m + l \\ && \hat a_k \leq z_k \leq \hat u_k, \hspace{0.1cm}k = 1, \cdots n  \\ && 0 \leq s_p < \infty, \hspace{0.3cm}p = 1, \cdots l \nonumber
\end{eqnarray}
where $ \hat e_i(\textbf{z, s}) = 0 $ is the combination of equality and inequality constraints in $ \textbf{P}_3 $. For simplicity of notation, we can use $ \textbf{y} = (\textbf{z}, \textbf{s}) $ to represent the variable vector, and $ \textbf{v} \leq \textbf{y} \leq \textbf{w} $ to denote the range of the variables. It should be noted that for some slack variables, the corresponding components of $ \textbf{w} $ can be set as infinite. 

As described before, the constraints are of nonlinear forms. To make the logic more clear and algorithm more understandable, we will first describe how to solve the linear constraints and then extend to the nonlinear forms of the constraints. Suppose the equality constraints in $ \textbf{P}_4 $ are in the linear form that $ {\textbf{E}}\textbf{y} = \textbf{c},  \textbf{y} \geq \textbf{0} $, where $ \textbf{E} \in \mathbb{R}^{(m+l) \times n}$ and $ \textbf{c} \in \mathbb{R}^{(m + l)\times 1}$. In addition, an nondegeneracy assumption is made here that every $ m + l $ columns of the matrix $ \textbf{E} $ are linearly independent and every basic solution to the constraints has at least $ m + l $ strictly positive values. This assumption can be easily satisfied since we can apply elementary transformation of matrix which will reduce the matrix be composed of independent columns or rows. With this assumption, every feasible point to the constraints will have at most $ n - m - l $ variables with values setting to zero. For any feasible point $ \textbf{y} $, it can be partitioned into two groups that $ \textbf{y} = (\textbf{y}_B, \textbf{y}_N) $, where $ \textbf{y}_B $ has the dimension $ m + l$ termed as basic variables, and $ \textbf{y}_N $ with dimension $ n - m - l $ is called as non-basic variable. Accordingly, matrix $ \textbf{E} $ can be decomposed as $ \textbf{E} = [\textbf{E}_B, \textbf{E}_N] $, where $ \textbf{E}_B $ and $ \textbf{E}_N $ are the columns corresponding to $ \textbf{y}_B $ and $ \textbf{y}_N $, respectively.

From the algebra we know that for each stage, the optimization of this problem is only dependent on the non-basic variables $ \textbf{y}_N $, since basic variable vector $ \textbf{y}_B $ can be uniquely determined from $ \textbf{y}_N $. A simple modification of the gradient decent method will provide a feasible improving direction $ \textbf{d} $ to optimize the objective function. A feasible improving direction $ \textbf{d} $ at the point $ \textbf{y} $ must follow:
\begin{equation}
\begin{split}
&\textbf{Ed} = \textbf{0}, \hspace{3.0cm} (a)
\\ &\nabla f(\textbf{y})^T\textbf{d} < 0,  \hspace{2.1cm} (b)
\end{split}
\end{equation}
where $ \nabla f(\textbf{y})^T $ is the gradient vector of objective function $ f(\textbf{y}) $ at point $ \textbf{y} $. Eq. (28a) means that if a feasible point $ \textbf{y} $ moves along the direction $ \textbf{d} $, the feasibility of the constraints will not be damaged. Eq.(28b) indicates that moving along $ \textbf{d} $ will make the objective function $ f(\textbf{y}) $ approach the optimal point. The \textit{reduced gradient} method as the following will find such moving direction $ \textbf{d} $ that satisfies Eq.(28).

The gradient vector corresponding to $ \textbf{y}_N $ (also called as \textit{reduced gradient}) can be found by the following expression:
\begin{equation}
\textbf{r}^T = \nabla_Nf(\textbf{y})^T - \nabla_Bf(\textbf{y})^T\textbf{E}_B^{-1}\textbf{E}_N
\end{equation}
where $ \nabla_Nf(\textbf{y})^T $ is the gradient vector of $ \nabla f(\textbf{y})^T $ that corresponds to $ \textbf{y}_N $ and $ \nabla_Bf(\textbf{y})^T $ is the gradient vector corresponding to  $ \textbf{y}_B $. From above reduced gradient, we can construct the feasible moving direction $ \textbf{d}_N $ that will move $ \textbf{y}_N + \lambda \textbf{d}_N $ in the feasible working space, where $ \textbf{d}_N  $ can be determined as the following:
\begin{equation}
d_{Ni}=\left\{
\begin{aligned}
&-r_i  \hspace{0.85cm} r_i \leq 0 \\
&-y_{Ni}r_i  \hspace{0.3cm}   r_i > 0 \\
\end{aligned}
\right.
\end{equation}
where $ d_{Ni} $ is the $ i $th element of $ \textbf{d}_N $, $ r_i $ is the $ i $th element of $ \textbf{r}^T $, and $ y_{Ni} $ is the $ i $th element of $ \textbf{y}_N $. Eq. (30) provides the rules for finding improving feasible direction for non-basic variables $ \textbf{y}_N $. Once the improving feasible direction for $ \textbf{y}_N $ is determined, we can get the corresponding moving direction $ \textbf{d}_B $ for $ \textbf{y}_B $ by expanding Eq.(28a):
\begin{equation}
\begin{split}
&\textbf{E}_N\textbf{d}_N + \textbf{E}_B\textbf{d}_B = 0 \\
&\textbf{d}_B = -\textbf{E}_B^{-1}\textbf{E}_N\textbf{d}_N
\end{split}
\end{equation}
Eq. (31) shows that $ \textbf{d}_B $ can be uniquely calculated from $ \textbf{d}_N $, and the moving direction is composed that $ \textbf{d} = [\textbf{d}_B, \textbf{d}_N] $. It can be proved that $ \textbf{d} = [\textbf{d}_B, \textbf{d}_N] $ satisfies Eq.(28a) and Eq.(28b), indicating both feasibility and improvability will be achieved for $ \textbf{d} $. 

The reduced gradient method dealing with the linear constraints can be generalized and extended to address the nonlinear constraints. Similar to linear constraints, we first partition the variables into basic and non-basic variable vector as $ \textbf{y} = (\textbf{y}_B, \textbf{y}_N) $, and the corresponding Jacobi matrix of $ \hat{\textbf{e}}(\textbf{y}) $ in $ \textbf{P}_4 $ can also be grouped into:
\begin{equation}
\dfrac{\partial \hat{\textbf{e}}}{\partial \textbf{y}} = (\dfrac{\partial \hat{\textbf{e}}}{\partial \textbf{y}_B}, \dfrac{\partial \hat{\textbf{e}}}{\partial \textbf{y}_N})
\end{equation}
and a nondegeneracy assumption is made here that for any point $ \textbf{y} $, $ \dfrac{\partial \hat{\textbf{e}}}{\partial \textbf{y}_B} \in \mathbb{R}^{(m + l)\times (m + l)}$ is nonsingular. 
\\ \indent For the case of nonlinear constraints, the \textit{reduced gradient} $ \textbf{r}^T $ with respect to $ \textbf{y}_N $ is expressed as:
\begin{equation}
\textbf{r}^T = \nabla_Nf(\textbf{y})^T - \nabla_Bf(\textbf{y})^T(\dfrac{\partial \hat{\textbf{e}}}{\partial \textbf{y}_B})^{-1}\dfrac{\partial \hat{\textbf{e}}}{\partial \textbf{y}_N}
\end{equation}
Now we specify the direction $ \textbf{d}_N $ as follows:
\begin{equation}
d_{Ni}=\left\{
\begin{aligned}
&0  \hspace{1.5cm} y_{Ni}r_i > 0 \hspace{0.2cm} \text{and} \hspace{0.2cm}y_{Ni} = v_i\\
&0 \hspace{1.5cm}  y_{Ni}r_i < 0 \hspace{0.2cm} \text{and} \hspace{0.2cm}y_{Ni} = w_i\\
&-y_{Ni}r_i  \hspace{0.4cm}   \text{otherwise} \\
\end{aligned}
\right.
\end{equation}
where $ v_i $ is the lower bound of the variable  $ y_i $ and $ w_i $ is the upper bound of the variable $ y_i $. However, the difference with the linear form is that $ \textbf{y}_N $ moves a straight line along $ \textbf{d}_N $, the nonlinear form of the constraints requires $ \textbf{y}_N $ move nonlinearly to continuously walk in the feasible space formed by the constraints. To address this, we can first move $ \textbf{y}_N $ along the direction defined by Eq. (34), then a correction procedure is employed making $ \textbf{y}_N $ return to working space to satisfy the feasibility of the constraints. Once a tentative move along $ \textbf{d}_N $ is made, the following iterative method can be used for the correction. Supposing $ \textbf{y}_k $ is the current feasible point, we first move the non-basic variable vector $ \textbf{y}_{N(k + 1)} = \textbf{y}_{Nk} + \lambda \textbf{d}_{Nk} $, to return point $ \textbf{y} = (\textbf{y}_{Bk}, \textbf{y}_{N(k + 1)}) $ near $ \textbf{y}_k $ back to the constraint space, we can solve the following equation:
\begin{equation}
\hat {\textbf{e}}(\textbf{y}_{Bk}, \textbf{y}_{N(k + 1)}) = 0
\end{equation}
for $ \textbf{y}_{Bk} $ where $ \textbf{y}_{N(k + 1)} $ is fixed. This is done by the following iterative procedure, which is described in Algorithm 3 from line 11 to 14:
\begin{equation}
\textbf{y}_{Bk_{j+ 1}} = \textbf{y}_{Bk_j} - (\dfrac{\partial \hat {\textbf{e}}(\textbf{y}_{Bk_j}, \textbf{y}_{N(k + 1)})}{\partial \textbf{y}_{Bk}})^{-1}\hat {\textbf{e}}(\textbf{y}_{Bk_j}, \textbf{y}_{N(k + 1)})
\end{equation}
where $ \textbf{y}_{Bk_j} $ is the basic variable vector of $ \textbf{y}_B $ in the $j$th iteration step according to Eq. (36). When this iterative process produces a feasible point $ \textbf{y}_{B(k + 1)} $, we have to check if following conditions are satisfied:
\begin{equation}
\begin{split}
&f(\textbf{y}_{B(k + 1)}, \textbf{y}_{N(k + 1)}) < f(\textbf{y}_{Bk}, \textbf{y}_{Nk})\\
&\textbf{v} \leq \textbf{y}_{B(k + 1)}, \textbf{y}_{N(k + 1)} \leq \textbf{w}
\end{split}
\end{equation}
If Eq. (37) holds true, it indicates that the new point is feasible and improvable. Then we set $ \textbf{y}_{k + 1}  = (\textbf{y}_{B(k + 1)}, \textbf{y}_{N(k + 1)})$ as a new approaching point, otherwise we will decrease the step length $ \lambda $ when we first make the tentative move for $ \textbf{y}_{Nk} $ and repeat the above iterative process. The procedure of generalized reduced gradient method is summarized as \textit{Algorithm 3}. 

\begin{algorithm}[!h]   
	\caption{ Secure outsourcing scheme for large-scale NLPs }   
	\label{alg:Framwork}   
	\begin{algorithmic}[1] 
		\REQUIRE ~~\\ 
		Starting point $ \textbf{y}_0 $ that $ \hat {\textbf{e}}(\textbf{y}_0) = 0 $; \\
		\ENSURE ~~\\ 
		Optimal result $\textbf{y}^*$ for $\textbf{P}_4$;\\  
		\STATE Initialize $k = 0$;   
		\STATE Decompose $ \textbf{y}_{0}  = (\textbf{y}_{B0}, \textbf{y}_{N0})$;
		\STATE Calculate Jacobi matrix of $ \hat {\textbf{e}}(\textbf{y}_0) $ and decompose the Jacobi matrix $ \dfrac{\partial \hat{\textbf{e}}}{\partial \textbf{y}} = (\dfrac{\partial \hat{\textbf{e}}}{\partial \textbf{y}_B}, \dfrac{\partial \hat{\textbf{e}}}{\partial \textbf{y}_N}) $ corresponding to $ (\textbf{y}_{B0}, \textbf{y}_{N0}) $;
		\STATE Compute $ \textbf{d}_{N0} $ from Eq. (34);
		\STATE \hspace{0.4cm} \textbf{while} $ \parallel \textbf{d}_{Nk} \parallel_1> \epsilon $
		\STATE \hspace{0.8cm} Choose $ \lambda > 0$ and compute $ \textbf{y}_{N(k + 1)} = \textbf{y}_{Nk} + \lambda \textbf{d}_{Nk} $;
		\STATE \hspace{0.76cm} If not $ \textbf{v} \leq  \textbf{y}_{N(k + 1)} \leq \textbf{w} $ :\\ \hspace{1.5cm} $ \lambda = 1/2 \lambda$;
		\STATE \hspace{1.5cm} go to (6) and repeat;
		\STATE \hspace{0.8cm} Initialize j = 0;
		\STATE \hspace{0.8cm} Let $ {\textbf{y}}_{Bj} $ = $ \textbf{y}_{Bk} $;
		\STATE \hspace{0.8cm} \textbf{while} $ \parallel \hat {\textbf{e}}(\textbf{y}_{Bj}, \textbf{y}_{N(k + 1)}) \parallel_1 > \epsilon$
		\STATE \hspace{1.1cm} Let $ E = (\dfrac{\partial \hat {\textbf{e}}(\textbf{y}_{Bj}, \textbf{y}_{N(k + 1)})}{\partial \textbf{y}_{Bk}})^{-1} $;
		\STATE \hspace{1.1cm} $ \textbf{y}_{B(j+ 1)} = \textbf{y}_{Bj} - E\hat {\textbf{e}}(\textbf{y}_{Bj}, \textbf{y}_{N(k + 1)}) $;
		\STATE \hspace{1.1cm} $ j = j + 1 $;
		\STATE \hspace{0.8cm} end \textbf{while};
		\STATE \hspace{0.8cm} If Eq. (37) holds true:
		\STATE \hspace{1.2cm} $ \textbf{y}_{B(k+1)} =  \textbf{y}_{Bj}$ ;
		\STATE \hspace{1.2cm} $ \textbf{y}_{k + 1}  = (\textbf{y}_{B(k + 1)}, \textbf{y}_{N(k + 1)})$;
		\STATE \hspace{0.8cm} else:
		\STATE \hspace{1.2cm} go to (6) and repeat;
		\STATE \hspace{0.8cm} $ k = k + 1 $;
		\STATE \hspace{0.8cm} Calculate $ \textbf{d}_{Nk} $ from Eq. (34);
		\STATE \hspace{0.5cm} end \textbf{\textbf{while}};  
		\STATE Let $ \textbf{y}^*= (\textbf{y}_{Bk}, \textbf{y}_{Nk})$;
		\RETURN $\textbf{y}^* $; 
	\end{algorithmic}  
\end{algorithm} 

Regrading verification of the correctness of the returned result, the users can apply KKT conditions of $ \textbf{P}_3 $ [8] to check if the return result is valid or not. 

\section{Performance Evaluation}\label{sec:evaluation}
In this section, we evaluate the performance of our proposed secure outsourcing protocol for large-scale NLPs with nonlinear constraints. For the experimental setup, the client side is implemented on a computer with Intel(R) Core(TM) i5-5200 U CPU processor running at 2.2 GHz, 8GB memory. For the cloud side, the experiment is conducted on a computer with Intel(R) Core(TM) i7-4770 U CPU processor running at 3.40 GHz, 16GB memory. We implement the proposed protocol including both the client and cloud side processed in Python 2.7. We also ignore the communication latency between users and the cloud for this experiment, since the computation dominates running time as demonstrated by our experiments.

We randomly generate a set of test cases that cover the small and large sized NLPs with nonlinear constraints, where the number of variables is increased from 1000 to 16000. The objective function here is randomly generated second-degree polynomial function, and nonlinear constraints are randomly generated with first-degree polynomial functions for equality constraints and second-degree polynomial functions for inequality constraints, respectively. All these test cases are carefully designed so that there are feasible under corresponding nonlinear constraints.

For the experiments, we first solve the original NLP in the client side, then solve the encrypted NLP in cloud side. Table \ref{tb1} shows the experiment results, and each entry in this table represents the mean of 20 trials. As illustrated in this table, the size of original NLPs is reported in the first three columns.  Besides, several parameters are adopted to evaluate the performance of proposed protocol.  $T_{original}$ is defined as the time to solve original NLP by client side. The time to solve encrypted NLP is divided into time for the cloud server $T_{cloud}$ and time for the client $T_{client}$. $T_{cloud}$ is defined as the time that cloud used to operate the encrypted NLP by the cloud server. $T_{client}$ is the time cost to encrypt and decrypt the original NLP by the client. Furthermore, we propose to assess the practical efficiency by two metrics calculated from $T_{original}$, $T_{client}$ and  $T_{cloud}$. The \textit{speedup} is calculated as $\frac{T_{original}}{T_{client}}$, representing time savings for the client to outsource the NLP to the cloud using proposed protocol. The speedup is expected to be greater than 1, otherwise there is no necessity for the client to outsource NLP to the cloud server. The \textit{cloud efficiency} is measured as $\frac{T_{original}}{T_{cloud}}$, indicating the time savings enabled by the cloud. It is expected that the encryption of the problem should not introduce  great overhead for solving the large-scale NLP. Moreover, due to more powerful computation capabilities of cloud server, the cloud efficiency is expected be grater than 1.

It can be seen that from the Table \ref{tb1} that the encryption can be finished in a very short time by the client. For instance, the time consumption of the encryption for the problem with 16000 variables is only 166.68s. However, the time cost to find the optimal solution by the cloud server is much longer but reasonable, and increases rapidly with growing number of variables. As shown in the penultimate column of the table, the speedup of proposed protocol increases dramatically when the size of the problem gets larger. Hence, a substantial amount of time can be saved for the client by proposed protocol. For example, the speedup is 49.66 for the problem with 16000 variables, indicating 97.9\% of time is saved for the client. The cloud efficiency is shown in the last column, and it can be seen that the cloud efficiency increases with the increasing size of the problem, indicating the powerful computation capabilities of cloud server. Consequently, the experiment results demonstrate our proposed secure outsourcing protocol is practical and efficient.

\begin{table*}[h]
	\caption{Performance Evaluation}
	\centering
\resizebox{\textwidth}{!}{
	\begin{tabular}{|c|c|c|c|c|c|c|c|c|}
		\hline
		\multicolumn{4}{|c|}{\textbf{Test Cases}} & \textbf{Original NLP} & \multicolumn{2}{|c|}{\textbf{Encrypted NLP} }& \textbf{Speedup} & \textbf{Cloud Efficiency} \\
		\hline
		\# & \# variables & \# equality constraints & \# inequality constraints & $T_{original}$(sec) &  $T_{cloud}$(sec) & $T_{client}$(sec) & $\frac{T_{original}}{T_{client}}$ & $\frac{T_{original}}{T_{cloud}}$ \\
		\hline
		1 & 1000 & 300 & 300 & 3.17  & 2.28   & 0.09  &   35.22 & 1.39 \\
		\hline
		2 & 2000 & 600 & 600 & 27.61 & 18.66  &  0.72 &  38.34  &  1.48\\
		\hline
		3 & 4000 & 1200 & 1200 & 182.05  & 112.38   & 4.48  &  40.62  &  1.62\\
		\hline
		4 & 8000 & 2400 & 2400 & 1236.42  & 695.79  &  36.50 &  34.33  & 1.77 \\
		\hline
		5 & 12000 & 3600 & 3600 & 2777.04  &  1368.55  & 75.32  &  37.02  &  2.03 \\
		\hline
		6 & 16000 & 4800 & 4800 & 8245.51  &  3720.11  &  166.68 &  49.66   &  2.21\\
		\hline
	\end{tabular}}\label{tb1}
\end{table*}

\section{Conclusion}\label{sec:conclusion}
In this paper, for the first time, we design an efficient and practical protocol for securely outsourcing large-scale NLPs with nonlinear constraints. The transformation technique is applied to protect the sensitive input/output information. In addition, we adopt the generalized reduced gradient method to solve the transformed NLP. A set of large-scale simulations are performed to evaluate the performance of proposed mechanism, and results demonstrate its high practicality and efficiency. It is expected that the proposed protocol can not only be deployed independently, but also serves as a building block to solve more sophisticated problems in the real world.

\bibliographystyle{splncs03}
\bibliography{INFOCOM2018}

\end{document}